%% file: main.tex
\documentclass[sigconf]{acmart}

\usepackage[htt]{hyphenat}
\usepackage{graphicx}
\usepackage[subtle]{savetrees}

\copyrightyear{2022} 
\acmYear{2022} 
\setcopyright{acmcopyright}\acmConference[ICSE-NIER'22]{New Ideas and Emerging Results }{May 21--29, 2022}{Pittsburgh, PA, USA}
\acmBooktitle{New Ideas and Emerging Results (ICSE-NIER'22), May 21--29, 2022, Pittsburgh, PA, USA}
\acmPrice{15.00}
\acmDOI{10.1145/3510455.3512777}
\acmISBN{978-1-4503-9224-2/22/05}

\begin{document}
\title{A Case for Microservices Orchestration Using Workflow Engines}

\author{Anas Nadeem}
\email{anas.nadeem@ndsu.edu}
\orcid{0000-0001-8859-6176}
\affiliation{%
  \institution{North Dakota State University}
  \city{Fargo}
  \state{North Dakota}
  \country{USA}
}
\author{Muhammad Zubair Malik}
\email{zubair.malik@ndsu.edu}
\affiliation{%
  \institution{North Dakota State University}
  \city{Fargo}
  \state{North Dakota}
  \country{USA}
}

\input{abastract}
\keywords{microservices, orchestration, debugging, workflows, fault-tolerance}
\maketitle

%these are just placeholder headings might needs to be changed!
\input{introduction}

\input{methdology}
\input{related_work}
\input{Discussion}
\input{conclusion}

%reference
\bibliographystyle{ACM-Reference-Format}
\bibliography{references}

\end{document}

%% file: abastract.tex
\begin{abstract}
Microservices have become the de-facto software architecture for cloud-native applications. A contentious architectural decision in microservices is to compose them using choreography or orchestration. In choreography, every service works independently, whereas, in orchestration, there is a controller that coordinates service interactions. This paper makes a case for orchestration.  The promise of microservices is that each microservice can be independently developed,  deployed, tested, upgraded, and scaled. This makes them suitable for systems running on cloud infrastructures. However, microservice-based systems become complicated due to the complex interactions of various services, concurrent events, failing components, developers’ lack of global view, and configurations of the environment. This makes maintaining and debugging such systems very challenging. We hypothesize that orchestrated services are easier to debug and to test this we ported the largest publicly available microservices' benchmark TrainTicket~\cite{ZhouPXSJLD21}, which is implemented using choreography, to a fault-oblivious stateful workflow framework Temporal~\cite{temporal}. We report our experience in porting the code from traditional choreographed microservice architecture to one orchestrated by Temporal and present our initial findings of time to debug the 22 bugs present in the benchmark. Our findings suggest that an effort towards making a transition to orchestrated approach is worthwhile, making the ported code easier to debug.
\end{abstract}

%% file: introduction.tex
%\vspace{-7pt}
\section{Introduction}
Service-oriented applications are increasingly becoming cloud-native and are built as a collection of small, independent, and loosely coupled microservices~\cite{Francesco17,Heorhiadi16}. Large web companies such as Tencent, Uber, Netflix, and Airbnb are increasingly building their core business systems using microservice  architecture~\cite{ZhouPXSJLD21}. The promise of microservices is that each microservice can be independently developed, deployed, tested, upgraded, and scaled. This makes them suitable for systems running on cloud infrastructures. However, these benefits come at a cost and microservice-based systems become complicated due to complex interactions of various services, events happening concurrently, components failing, developers’ lack of global view, and configurations of the environment. This complexity and dynamism of microservice systems pose unique challenges for system developers and makes them hard to implement and debug.   

Choreography is traditionally the more prevalent way of organizing microservices~\cite{Mahtab21}. It follows the event-driven paradigm in which every service works independently. There are no hard dependencies between the microservices and they are loosely coupled only through shared events. Each service listens for the events that it is interested in, and is fully responsible to react to these events and perform its operations or business logic independently. This involves dealing with all types of failures. Frameworks such as Spring Boot~\cite{spring} makes it easy to create stand-alone microservices by embedding servelet capabilities, REST client capability, database integration, and externalized configuration in a standalone Spring application. Containers such as Docker~\cite{docker} make it easy to configure and deploy these applications. Organization and management of such services include discovery and load management which is enabled by frameworks such as Kubernetes~\cite{k8s}.    

Microservices are arranged together to implement workflows, which are repeatable patterns of activity that naturally arise from the systematic organization of resources and information. This happens organically in choreography and no one entity is responsible for end-to-end monitoring of system or business workflow. This makes implementing business logic easier but failure recovery logic becomes much more complicated. To ensure availability in case of infrastructure outages in the cloud, a microservice must guard itself against failures of its dependencies~\cite{Gremlin}.

 In contrast, a workflow engine makes it easier to explicitly orchestrate microservices by managing low-level distributed programming issues and enables the microservice to focus on implementing the business logic. Historically, distributed workflow engines have been implemented to execute complex, computationally intensive, and repetitive tasks such as gene sequencing in bio-informatics~\cite{bioinformatics}. However, these earlier workflow engines were very domain-specific, non-resilient, and non-programmable. An example of a programmable distributed framework is MapReduce~\cite{MapReduce}, as it provided a programming model that enabled data analysts to write distributed tasks without having to deal with low-level distributed programming details. However, MapReduce is not an engine in the sense that it does not control and manage the life-cycle of workflow and is non-resilient. Temporal~\cite{temporal} is a workflow engine for microservices that manages the workflow life cycle and also provides a fault-oblivious stateful workflow programming model to orchestrate microservices.
While various systems such as Netflix Conductor~\cite{conductor}, Uber Cadence~\cite{cadence}, Apache Airflow~\cite{apacheairflow} also provide microservice orchestration capabilities, our criterion for opting to use Temporal was for the following reasons \textbf{1)} ease of use as a result of workflow definition in plain code rather than a complex Domain-Specific Language \textbf{2)} framework support for debugging.

Using Temporal workflow engine, programmers write their code in a familiar language while the platform orchestrates the tasks and makes the application resilient to all failures. This means any state within the workflow is durable and frees up the developer to focus on the business logic of the application rather than spend most of the time building resilience. Code written for Temporal needs to adhere to certain constraints, for example, a purely random variable within part of the code that needs to re-execute, can break stateful-workflow assumptions: executions must be idempotent and deterministic. These are semantic errors that will break part of the code which otherwise will normally be considered valid. We implemented a simple linter~\footnote{\url{https://github.com/arise-ndsu/temporalint}}
to deal with such issues and our experience in porting code from original microservices bugs benchmark to temporal framework was fairly smooth.

We make the following contributions: \textbf{a)} We port the largest publicly available microservice benchmark to a fault-oblivious stateful workflow engine and report our experience; \textbf{b)} We experimentally evaluate debugging of 22 bugs present in the TrainTicket benchmark using debugging process supported by Temporal; \textbf{c)} We compare our results with experimental results on original benchmarks and present our observations and insights to provoke further research.    
The experiments are released as an open-source project~\footnote{\url{https://arise-ndsu.github.io/trainticketworkflows}} and can be replicated with a simple set-up.

%% file: methdology.tex
\section{Transitioning to Temporal}
Moving to the Temporal framework requires a paradigm shift to writing code in comparison to the traditional methods of developing microservice systems. While it hides the complexity of handling various edge cases resulting in a resilient and fault-tolerant system, transitioning to Temporal framework-based system requires some refactoring. The framework requires the separation of business logic from control logic. Subsequently, at a high level, every business process is accomplished through orchestrating a \emph{workflow}.
All the interaction between microservices must be managed by Temporal. Any non-deterministic action that is prone to failure must be wrapped by an \emph{activity}.
These activities and workflows are finally registered with a \emph{worker} which is responsible for picking these tasks from a queue within Temporal, and finally executing these tasks.
To better understand the idea of how to transition from a traditional interaction between several microservices to Temporal, we illustrate mapping one such interaction in the benchmark system and employ a similar strategy to map all the business processes within TrainTicket to \emph{workflows} and \emph{activities}. 

\subsection{Background: Benchmark System}
To systematically study how Temporal constructs; workflows, activities and workers, map to a realistic microservice system, we port TrainTicket~\cite{ZhouPXSJLD21} to Temporal. This system provides typical train ticket booking functionalities such as purchase, cancellation, and changing by making use of over 30 fine-grained microservices written in multiple languages including Java, NodeJS, Python, and Go. In addition to that, TrainTicket incorporates 22 fault cases collected from an extensive industrial survey in order to serve as a benchmark system for microservices-based research.  We deploy this benchmark system along with the Temporal server on a machine with a 20GB available memory, in the form of Docker containers. After mapping the benchmark system, we study the role of Temporal in 1) assisting the system cope up with and recover from these faults 2) enabling debugging practices towards fault localization and debugging.

\begin{figure}[t]
    \includegraphics[width=8cm]{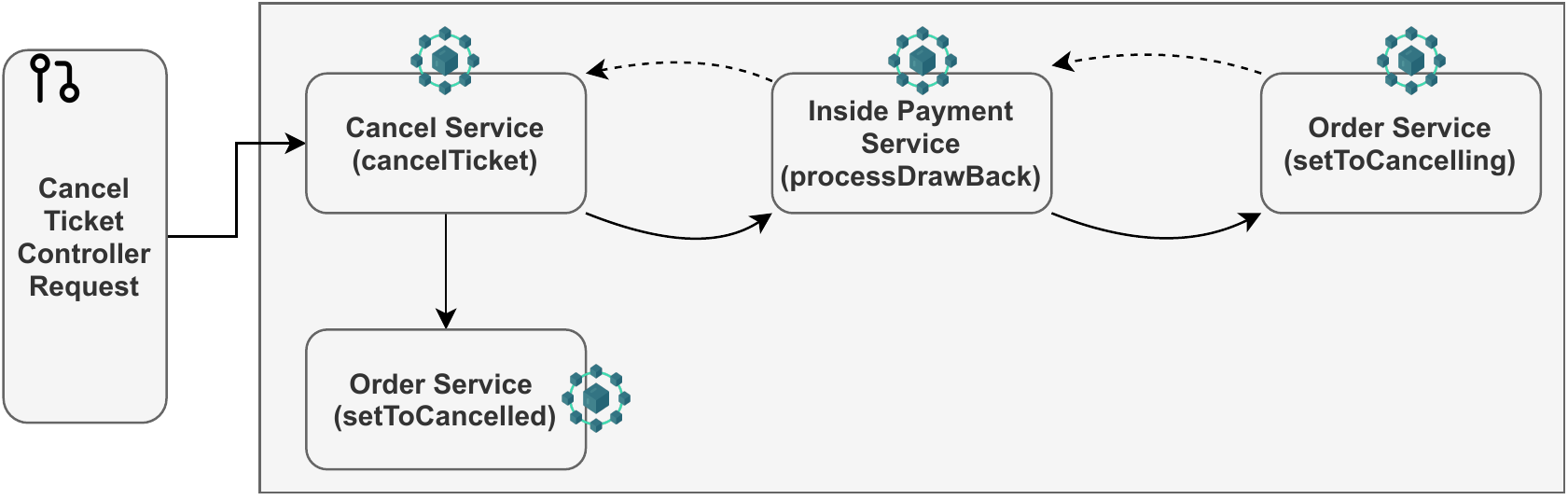}
    \vspace{-10pt}
    \caption{\small{original cancelTicket process within TrainTicket}}
    \label{canceloriginal}
\end{figure}

\subsection{Original Process}
One of the business processes in the benchmark system allows users to cancel any purchased tickets. This process involves two major events 1) refunding the user 2) changing the status of the order to cancelled. The holistic view of the services involved in this business process is illustrated in Figure~\ref{canceloriginal}. When a user requests the cancellation of an existing ticket, the \texttt{cancel-service} is invoked. The cancel-service then invokes \texttt{inside-payment-service} in order to refund the purchase amount back to the user. The same \texttt{inside-payment-service} then tells \texttt{order-service} to set the status of the order to cancelling, and refunds the money back to the user. Finally, the \texttt{cancel-service} itself calls the \texttt{order-service} again to mark the order as cancelled.

\begin{figure}[t]
    \includegraphics[width=8cm]{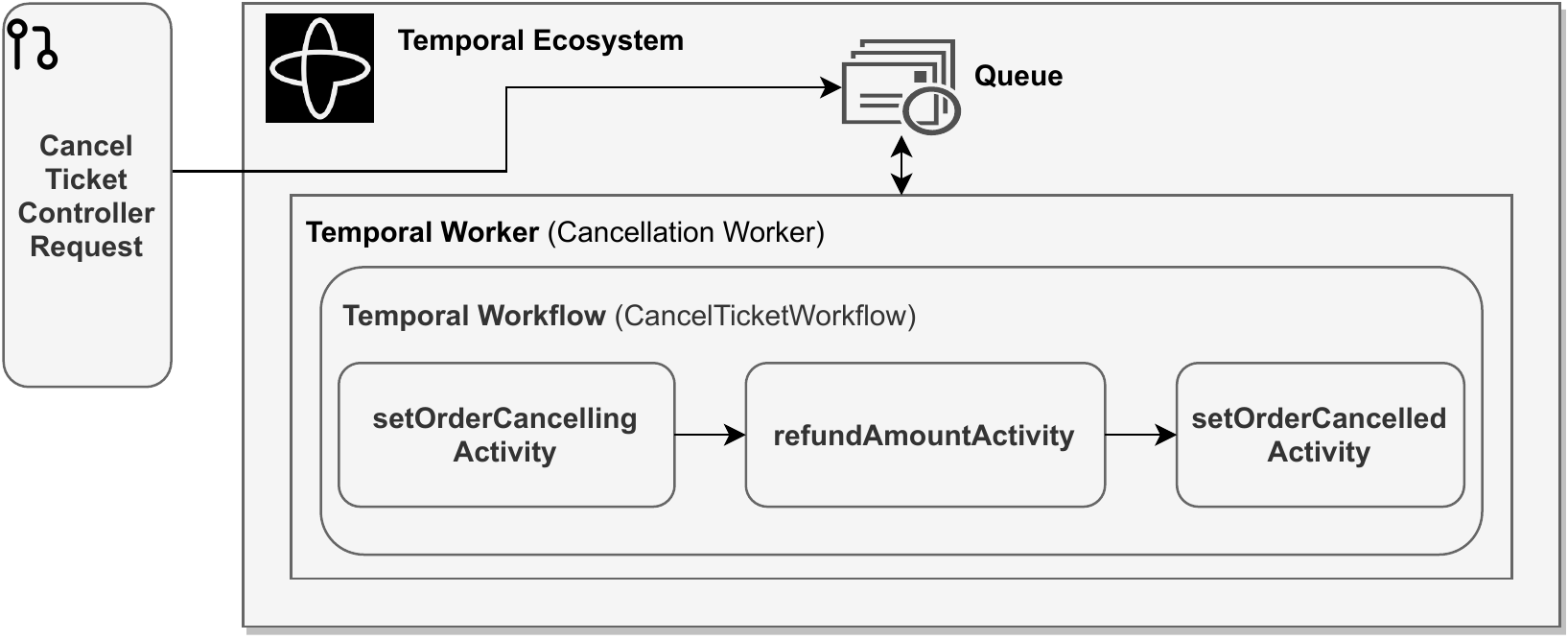}
    \vspace{-10pt}
    \caption{\small{cancelTicket workflow after porting to Temporal}}
    \label{cancelworkflow}
\end{figure}

\subsection{Process Expressed in the Temporal Way}

From a Temporal view point, the previously discussed business process can be systematically mapped into a \texttt{cancelTicket} Workflow. Any subsequent calls to any microservice required during the process of cancelling a ticket, must originate from within the \emph{activities} construct of Temporal. Therefore, we mapped the above call hierarchy to the following: 1) upon a cancel request, the controller triggers a \texttt{cancelTicket} workflow 2) the \texttt{cancelTicket} workflow first invokes \texttt{setOrderCancellingActivity} which invokes the \texttt{order-service} to set the status to cancelling 3) the \texttt{cancelTicket} Workflow then invokes \texttt{drawBackMoneyActivity} which internally uses the \texttt{inside-payment-service} to process the refund 4) the \texttt{cancelTicket} workflow finally calls \texttt{setOrderCancelledActivity} which invokes the \texttt{order-service} and sets the status of the order to \texttt{cancelled}.

Temporal system is event-driven in nature. Although we discussed the execution of our workflow as it was executing directly with the request, however, in reality, a request only registers an event into the Temporal queue. This event is then picked up by a Temporal worker and executed within the worker process. Hence, the worker construct serves as the baseline task distribution mechanism within Temporal. Therefore, we register all of our functional entities including the \texttt{cancelTicket} workflow, and all the activities with a \texttt{CancellationWorker}. Figure~\ref{cancelworkflow} illustrates the \texttt{cancelTicket} process mapped into Temporal constructs.

\section{Debugging Faults In Temporal}
To illustrate the debugging process within Temporal, we consider the same workflow that we discussed in the previous section and follow a similar strategy for the rest of the faults. We use the benchmark system with a fault injected within the \texttt{cancelTicket} workflow. To inject this fault case, TrainTicket slightly modifies the system to lack strict sequence control and simulate network congestion while invoking the \texttt{inside-payment-service} for the refund functionality and \texttt{order-service} to set the order status to \texttt{cancelled}. With a simulated network congestion, the refund through \texttt{inside-payment-service} gets triggered after the \texttt{order-service} has already set the status to cancelled, due to which \texttt{inside-payment-service} fails to further proceed with the request. After mapping the same workflow within, we invoke the \texttt{drawBackMoneyActivity} and \texttt{setOrderCancelledActivity} asynchronously to replicate the scenario within the \texttt{cancelTicket} workflow. We highlight the steps that were involved in debugging this fault within the Temporal ecosystem, in the context of a formal debugging model~\cite{Li2019}.

\begin{figure}[t]
        \includegraphics[width=7cm]{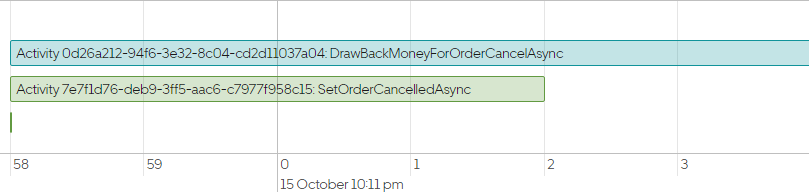}
        \vspace{-10pt}
        \caption{\small{Visual trace of faulty cancelTicket workflow on Temporal Web UI}}
        \label{visualtrace}
        \vspace{-10pt}
    \end{figure}

\begin{itemize}
    \item \textit{Problem Space Construction:}
    The problem space construction was a perception development step in the debugging process and involved developing an initial understanding of the fault. We observe that for the fault under consideration, Temporal Web UI~\cite{temporalwebui} produces a stack trace, as well as a visual trace graph, illustrated in Figure~\ref{stacktrace} and Figure~\ref{visualtrace}, respectively. We collect and observe these logs and traces to develop a preliminary understanding of the fault.
    \begin{figure}[t]
        \includegraphics[width=9cm]{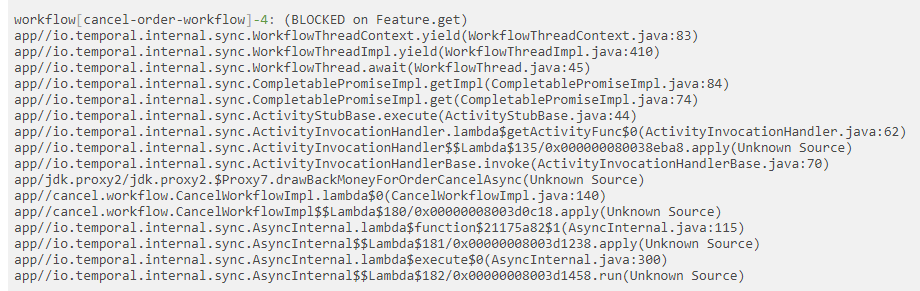}
        \vspace{-18pt}
        \caption{\small{Stack trace of faulty cancelTicket workflow on Temporal Web UI}}
         %\caption{\small{Stack trace of the fault on Temporal Web UI}}
        \label{stacktrace}
        \vspace{-10pt}
    \end{figure}

    \item \textit{Identification of Fault Symptoms:}
    This step involved setting up an environment to reproduce the fault. Since Temporal makes business processes explicit in terms of a workflow, we quickly identified the microservices involved in the workflow. Moreover, this straightforwardness of the workflow also helped us speed up the environment setup by enabling us to create a minimalist version of the environment to include only the services participating in the workflow.
    After the setup, we execute the fault case to evaluate the system outputs and compare them with the expected outputs to identify discrepancies.
    \item \textit{Fault Diagnosis:}
    In this step, we use our knowledge gained in the prior steps to hypothesize the location of the fault. Through thorough analysis using the debugging functionality on Temporal Web UI, we identify the root cause and the precise location of the fault. With the help of the holistic trace view of the workflow on Temporal from its invocation to where it incurred the fault, we reached the exact location within the \texttt{cancelTicket} workflow which triggered the fault. Analyzing the visual trace in Figure~\ref{visualtrace}, we observe that \texttt{setOrderCancelledActivity} completes prior to \texttt{drawBackMoneyActivity} which breaks the intended expectations.
    \item \textit{Solution Generation and Verification:}
    The final steps involved fixing the identified fault. The fault was corrected by adding the necessary sequence control after which the system automatically resumed its execution and resulted in the completion of the workflow. Final steps involved the execution of necessary tests in order to verify that the fault was resolved.
    
\end{itemize}

\input{Tables/fault_debug_times}

%\noindent\textbf{Testing and Debugging:}
%Temporal's SDK provides a test framework to facilitate testing Workflow implementations. The framework supports unit tests as well as functional tests of the Workflow logic. Temporal provides a Web UI that can be used to view Workflow Execution states or explore and debug Workflow Executions. Each workflow can be explored based on its ID and argument, variables and other parts of state and history are also fully available. There is no need to use any external tool for debugging. In comparison to the original microservices benchmark paper had to use external tools in their empirical study of debugging process.
\vspace{-8pt}
\subsection{Preliminary Findings}
%Table~\ref{debuggingtimes} sums up our findings from the debugging of the fault-cases replicated from the benchmark system along with the overall time (in hours) required to debug each fault in the original TrainTicket study and compare them with the times that were taken by a developer; with more than two years of industry experience with microservices, to debug these faults after mapping all these faults in Temporal workflows, who had no prior knowledge of the faults. We found out that by utilizing a mixture of debugging capabilities offered by Temporal, the time required to debug these fault cases was lesser than the overall time taken in the original study. We also found that Temporal self-contains necessary tools to support debugging and testing of the system, in comparison to the microservice-benchmark paper, which employed external tools for the debugging process.

Table~\ref{debuggingtimes} sums up our preliminary findings from a user case study. The table reports the debugging time by our user, of the replicated fault-cases from the benchmark system against the time required to debug the same fault in the original TrainTicket study. User in our study had more than two years of industry experience with microservices development and had no prior knowledge of the faults. However, the user went through elementary training of available debugging support on Temporal. We note that by utilizing the mixture of debugging capabilities offered by Temporal, the time required to debug these fault cases was lesser than the overall time taken in the original study. We also found that Temporal self-contains necessary tools to support debugging and testing of the system, in comparison to the microservice-benchmark paper, which had to employ external tools for the debugging process.

Consistent with the fault types in the original benchmark system, we discuss how Temporal responded to each category of faults. The faults in the benchmark system broadly fall into the categories: \textbf{Internal} faults are caused as a result of implementation in the microservice itself. \textbf{Interaction} faults are caused as a result of interactions between various microservices. \textbf{Environment} faults occur as a result of misconfiguration of the infrastructure~\cite{ZhouPXSJLD21}.

Temporal plays no significant role in the debugging of faults that are internal in nature, as the cause of these faults mainly lies in misinterpreted requirements and typically do not cause any failure within the system. In the case of faults in interaction of microservices, Temporal fully preserved the state of the execution. The execution was resumed by Temporal automatically as soon as the bug was resolved. Similarly, upon the occurrence of an environment fault due to a service downtime or service unavailability, Temporal managed to fully preserve the current execution state and resumed execution as soon as the unavailable microservice came back up.

%% file: Tables/fault_debug_times.tex
% \usepackage{graphicx}
\begin{table*}[]
\centering
\resizebox{0.9\textwidth}{!}{%
\begin{tabular}{|l|l|l|l|l|}
\hline
\textbf{Fault} & \textbf{Description} & \textbf{Type} & \textbf{Overall Time (H)} & \textbf{Overall Time Temporal (H)} \\ \hline
Fl & Lack of sequence control in   asynchronus message delivery in cancelTicket process & Interaction & 13.6 & 2.4 \\ \hline
F2 & Network congestion in ticket   reservation causes delivers requests in the wrong order & Interaction & 13.9 & 3.6 \\ \hline
F3 & Requests occupy larger memory   than the available resources causing service unavailability & Environment & Failed & 0.5 \\ \hline
F4 & SSL offloading in each   microservice causes prolonged response time & Environment & Failed & 0.4 \\ \hline
F5 & Incoming requests to basic info   service exceed the available threadpool size causing a timeout & Interaction & 12.6 & 5.2 \\ \hline
F6 & Recursive errors in voucher   service results in a large number of retries leading to a time out & Interaction & 5.9 & 6 \\ \hline
F7 & Call to charge amount during the   ticket purchase process intermittently times out & Interaction & 12 & 5.3 \\ \hline
F8 & Missing request tokens during   the ticket reservation process for VIP users leads to failure & Interaction & 12.2 & 4.8 \\ \hline
F9 & Words on UI have incorrect   display alignment & Internal & 1.8 & 0.9 \\ \hline
F10 & Ticket reservation process makes   incorrect API calls resulting in failure & Interaction  & 10.6 & 6.1 \\ \hline
F11 & A missing edge case causes   intermittent lack of sequence control during the cancellation process   resulting in failure & Interaction & 13.9 & 4.8 \\ \hline
F12 & A cancellation request to the   order service for a locked station rejects any incoming requests resulting in   failure & Interaction & 19.3 & 8.1 \\ \hline
F13 & A transmission delay in   simultaneous requests by the same user over a short period of time puts the   system in an inconsistent state & Interaction & 16 & 7.6 \\ \hline
F14 & Calculation of the price of a   seat is wrong & Internal & 2.9 & 3 \\ \hline
F15 & A lengthy request body size   results in nginx to block the request & Environment & 1.8 & 0.6 \\ \hline
F16 & Adding routes by file upload   with a size bigger than the limit results in rejection of the file & Environment & 5.9 & 6.1 \\ \hline
F17 & Requesting voucher with a   simulated load delay in sql results in a query timeout & Internal & 5.9 & 4.1 \\ \hline
F18 & A missing null value check   during the train selection process results in an error in getFood response & Internal & 3.4 & 3.9 \\ \hline
F19 & Display of package consignment   prices in French is in wrong format & Internal & 0.7 & 0.7 \\ \hline
F20 & Mismatch in version of a common   library versioning results in loading different versions of same data   structure & Environment & 3.8 & 3.1 \\ \hline
F21 & Missing aria-labeled-by in   verification code field in login results in poor accessibility & Internal & 1.6 & 1.5 \\ \hline
F22 & A mismatch of column name in the   select and from part of an sql query results in empty results during voucher   printing & Internal & 0.4 & 0.5 \\ \hline
\end{tabular}%
}
\caption{\small{Comparison of debugging times between original study and system mapped into Temporal}}
\label{debuggingtimes}
\vspace{-20pt}
\end{table*}

%% file: related_work.tex
\vspace{-6pt}
\section{Related Work}
Microservices is one of the fastest-growing areas in software engineering~\cite{marketcap}, however, there is limited high-quality research being done in the field. Hassan et al.~\cite{designTradeOff} considered design trade-offs for microservices along the dimensions of size/number and the global/local non-functional requirement satisfaction. More recent work~\cite{Mahtab21} has focused on qualitative analysis of composing microservices by choreography or orchestration. However, to the best of our knowledge, we are the first to present a quantitative comparison of compositional choices on a single benchmark. Zhou et al.~\cite{microBenchmarks} provide a good literature review of publicly available benchmarks of microservices and informed our choice for TrainTicket. Debugging microservices has gained much attention recently. Our work is highly inspired by a detailed survey, benchmark and empirical study by Zhou et al.~\cite{ZhouPXSJLD21}. They recorded experiences and processes of practitioners of varied skill and experience levels, systematically recreated the most common bugs experienced by practitioners and performed a detailed study of commonly used debugging techniques and effort required for each one of them. This became the basis of our work and comparison. Heorhiadi et al.~\cite{Gremlin} provided a framework for systematic testing of failure handling capabilities of microservices. Service call graphs, service usage logs, and attribute graphs have been used recently~\cite{microRCA,microHECL} for root cause analysis of anomaly detection and debugging of availability issues. Temporal provides highly detailed logs and call histories that can benefit from such analysis.
%Related work can be explored from three perspectives a) Choice Between 

%% file: Discussion.tex
\vspace{-10pt}
\section{Discussion}
\textbf{Choreography: Promise vs Reality:} 
Loose service coupling and strong cohesion in a choreographed microservices architecture seem very promising for agility and fault tolerance. It makes adding and removing services as simple as connecting or disconnecting a service from an appropriate channel in the event broker. Loose coupling also implies that choreography isolates microservices, such that if one service fails, other services not dependent on it can carry on while the issue is rectified. Choreographed, event-driven microservices allow for development teams to operate more independently and focus on their key services. Once these services have been created, they are now easily able to be shared between teams. 
However, teams that have built larger systems learn that much technical debt is incurred in the process~\cite{conductor}. Process flows are “embedded” within the code of multiple services. This replication of code is a headache for maintenance. Also,  there are tight coupling and assumptions around input/output and other service level agreements that make it harder to adapt to changing needs. Further, many critical system-level questions cannot be answered immediately, such as:  “How much are we done with process X”?

\noindent\textbf{A Lesson from History of Computing:}
In the early days of information systems, practically everybody wrote their own data store and spent a significant portion of their project time managing data management code, which was cumbersome and error-prone~\cite{redbook}. It was followed by standardized libraries and hierarchical systems until Codd presented the relational model of data~\cite{Codd70}, and Chamberlin and Boyce introduced \emph{A Structured English Query Language}~\cite{SEQUEL}. We envision that workflow engines will do for microservices what Relational Database Systems did for information systems. We argue that workflow engines can relieve developers from focusing on low-level distributed programming concerns (such as implementing ACID constraints) and enable them to focus on implementing business logic.

%********************ideas for conclusion***********************

\noindent\textbf{Fault-oblivious Stateful Programming:}
Temporal Server handles the durability, availability, and scalability of the application. In terms of CAP theorem~\cite{CAP}, each server instance is eventually available and highly consistent. In effect, Temporal Server provides a durable virtual memory per workflow execution, that is not linked to any specific process. It preserves the full application state (including program stacks with local variables) across all kinds of software and hardware-related failures. Temporal SDK builds on these capabilities and enables users to write their application code using the full power of the programming language. Temporal ensures that a triggered call will never fail, thus leading to the utility of long-running code spanning over multiple days or even months (let's say a method that needs to perform something after every 7 days).

%\noindent\textbf{Testing and Debugging:}
%Temporal's SDK provides a test framework to facilitate testing Workflow implementations. The framework supports unit tests as well as functional tests of the Workflow logic. %Temporal provides a Web UI that can be used to view Workflow Execution states or explore and debug Workflow Executions. Each workflow can be explored based on its ID and argument, variables and other parts of state and history are also fully available. There is no need to use any external tool for debugging. In comparison to the original microservices-benchmark paper had to use external tools in their empirical study of debugging process.

%***********************************

%% file: conclusion.tex
\vspace{-1pt}
\section{Conclusion}
In this work, we have ported the TrainTicket benchmark reflecting typical faults of a microservice to a novel fault-oblivious stateful workflow orchestration engine Temporal. We have used the replicated faults to assess the time to debug the faults in orchestration-based microservice implementation using Temporal’s Web UI that supports stack tracing and visual tracing. We observed that orchestration makes debugging easier and faster. In the future, we plan to run the presented study with more participants, port other benchmarks to Temporal, and study its performance on availability issues.

\section{Acknowledgements}
We are extremely grateful to Mr. Samar Abbas for his insightful discussions and technical support for Temporal.

\begin{comment}
We acknowledge the support of Mr. Samar Abbas for making this study possible.
\end{comment}